\documentclass{PoS}

\usepackage[utf8]{inputenc}
\usepackage{mathrsfs}
\usepackage{amssymb}
\usepackage{amsmath}
\usepackage{latexsym}
\usepackage{graphicx}

      \newcommand{\be}{\begin{equation}}
      \newcommand{\ee}{\end{equation}}  	 
      \newcommand{\ba}{\begin{eqnarray}}      
      \newcommand{\ea}{\end{eqnarray}}	 
      \newcommand{\bite}{\begin{itemize}}
      \newcommand{\eite}{\end{itemize}}

\title{The SF running coupling with four flavours of staggered quarks\thanks{FT-UAM-07-16, IFT-UAM-CSIC-07-49,TCDMATH-07-13}}

\ShortTitle{SF with 4 flavours of staggered quarks}

\author{\speaker{Paula P\'erez Rubio}\thanks{Support by the Spanish Ministery of Education and Science under grant FPU}\\
        Universidad Aut\'onoma de Madrid, Instituto de F\'{\i}sica Te\'orica, 28049 Cantoblanco, Spain\\
        and
	Trinity College Dublin, 
        School of Mathematics, Dublin 2, Ireland\\
        E-mail: \email{paula.perez@uam.es}}

\author{Stefan Sint\\
	Trinity College Dublin, School of Mathematics, Dublin 2, Ireland\\
        E-mail: \email{sint@maths.tcd.ie}}

\abstract{In order to study the running coupling in four-flavour QCD, we review the set-up of the Schr\"odinger functional (SF) with staggered quarks. 
Staggered quarks require lattices which, in the usual counting, have even spatial lattice extent $L/a$ while the time extent $T/a$ must be odd. Setting $T=L$ is therefore only possible up to ${\rm O}(a)$, which introduces different cutoff effects already in the pure gauge theory. We re-define the SF such as to cope with this situation and determine the corresponding classical background field. A perturbative calculation yields the coefficient of the pure gauge ${\rm O}(a)$ boundary counterterm to one-loop order.}

\FullConference{The XXV International Symposium on Lattice Field Theory\\
		 July 30-4 August 2007\\
		 Regensburg, Germany}

\begin{document}
\section{Introduction}

The renormalised coupling in the Schr\"odinger functional (SF) scheme has been 
defined in~\cite{Luscher:1992an,Sint:1995ch} and its scale evolution has been studied in QCD with zero and two quark flavours~\cite{Luscher:1993gh,Della Morte:2005kg}. We here discuss the set-up for studying the running coupling with four quark flavours, in the lattice regularisation with staggered 
quarks~\cite{Miyazaki:1994nu,Heller:1997pn}. This writeup is organized as follows.  We start with  a reminder of the basic definition of the renormalised coupling in a formal continuum notation. Its lattice formulation with staggered quarks will require a modification at O($a$) of the standard framework, even in the pure gauge theory.The corresponding tree-level and one-loop computations are then described and we end with an outlook to future work.

\section{The renormalised Schr\"odinger functional coupling}

The Schr\"odinger functional is a useful tool 
to study the scaling properties of QCD. It is defined as the Euclidean time evolution kernel for a state at time $x_0 = 0$ to another state at Euclidean time $x_0 = T$.  Using the transfer matrix formalism, it can be written 
as a path integral over fields which satisfy Dirichlet boundary conditions at Euclidean times $x_0 = 0$ and $T$ and $L$-periodic boundary conditions in space.
More precisely, one imposes homogeneous boundary conditions on the quark fields,
\ba
(1 + \gamma_0)\psi \big|_{x_0 = 0} &= 0 = &  (1 - \gamma_0 )\psi\big|_{x_0 = T} \nonumber\\
\bar \psi (1 - \gamma_0)\big|_{x_0 = 0} &= 0 =  & \bar \psi (1 + \gamma_0 )\big|_{x_0 = T},
\ea
while the spatial components of the gluon fields satisfy
\be
A_k(x) \big|_{x_0 = 0} = C_k \qquad A_k(x)\big|_{x_0 = T } = C'_k.
\label{bc_cont}
\ee
The SF is considered a functional of these boundary gauge fields,
\be
\mathcal Z[C', C] = \int \mathcal D[ A, \psi, \bar \psi] e^{-S[A, \psi, \bar \psi]}.
\ee
To define the SF coupling we follow\cite{Luscher:1992an}
and choose Abelian and spatially constant gauge boundary fields $C$ and $C'$,
\be\label{BC}
C_k = \frac{i}{L} {\rm diag }(\phi_1, \phi_2, \phi_3),\qquad
C'_k = \frac{i}{L} {\rm diag }(\phi'_1, \phi'_2, \phi'_3).
\ee
Under mild conditions on these angular variables, the absolute
minimum of the action is uniquely determined up to gauge equivalence.
Denoting this minimal action configuration by $B_\mu(x)$,
one may thus unambiguously define the effective action
\be\label{Effaction}
\Gamma[B] = - \ln \mathcal Z[C, C'].
\ee
In the weak coupling domain, the SF can be computed by performing a saddle point expansion of the functional integral about the induced background field $B$,
leading to an asymptotic series of the form,
\be
\Gamma[B]\stackrel{g_0\to 0}{\longrightarrow} \frac{1}{g_0^2}\Gamma_0[B] + \Gamma_1[B] + g_0^2 \Gamma_2[B] + \dots.
\ee
The leading term of the series is given by the classical action, while the higher order contributions are sums of vacuum bubble Feynman diagrams with an increasing number of loops. 

The finite size scaling technique is based on the idea of a renormalized coupling that runs with the box size $L$. In order for $L$ to be the
only external scale in the system all dimensionful parameters are taken
proportional to $L$. In particular, one sets $T=L$ and the quark masses 
are set to zero. Then one lets the background field $B$ depend
on a dimensionless parameter $\eta$ and defines a renormalized coupling,
\be
\frac{1}{\bar g ^2(L)} = \frac{\Gamma'_0}{\Gamma'},
\ee
where the prime indicates differentiation with respect to $\eta$ at $\eta=0$,
\be
\Gamma' = \frac{\partial}{\partial\eta} \Gamma\left[B(\eta)\right]\big|_{\eta = 0}.
\ee
Note that the numerator $\Gamma_0'$ merely serves as a normalization constant
to ensure $\bar g = g_0$ at tree level. When initially defined with the lattice regularisation in place the coupling is well-defined beyond perturbation theory,
In particular, its non-perturbative running
has been previously computed for a pure Yang Mills theory in \cite{Luscher:1993gh} and for QCD with two flavours in \cite{Della Morte:2005kg}. Here, we prepare the set-up for similar studies in QCD with 4 flavours. Staggered fermions seem to be a natural choice, as they come in multiples of 4 species, due to the incomplete elimination of the doubler modes.

\section{Subtleties with staggered fermions}

\begin{figure}[ht]
\centerline{
\includegraphics{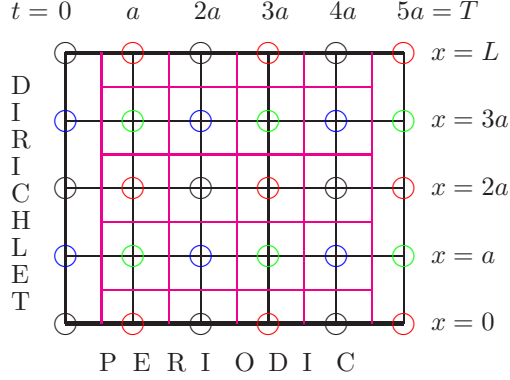}
}
\vspace{-0.0cm}
\caption{$2^d$ fermionic degrees of freedom on a $d=2$ dimensional lattice.\label{figure2}}
\end{figure}

As observed previously in~\cite{Miyazaki:1994nu,Heller:1997pn}, the SF
with staggered fermions requires lattices where the time extent $T/a$ is odd 
while the spatial extent $L/a$ has to be even. This is illustrated in Fig.~\ref{figure2}. As the Dirac spinors are
reconstructed from the one-component fields on the corners of a hypercube 
(indicated by different colours) the constraint arises from having the degrees of freedom for a multiple of 4 Dirac spinors fit on the lattice, taking into account that half the Dirac spinors satisfy Dirichlet conditions at the time boundaries.

The continuum limit for the SF coupling is usually taken setting $T=L$
already at finite values of the lattice spacing. Obviously, with staggered fermions this can only be done up to terms of O($a$). In order to define the continuum limit, it is convenient to imagine a dual lattice, as indicated in Fig.\ref{figure2}. For the spatial directions, the dual lattice has the same number of points as the original lattice. However, in the time direction, the number of points is reduced by one. Denoting the temporal length of the dual lattice by $T'/a$  one then has $T' = L$. 
A further justification may be obtained by looking at the fermionic degrees of freedom: the four Dirac spinors are reconstructed from the $2^4$ one-component spinors of the corners of a hypercube and may be assumed to live on a lattice with nodes in the center of each hypercube. This is nothing but the dual lattice where every second node is eliminated in each direction. In order to get an integer multiple of 4 Dirac spinor fields, it is thus necessary that both $L/a$ and $T'/a$ are even. Finally, the case $T = L-a$ can be treated the same way, if one imagines a modified dual lattice overlapping by $a/2$ in both time directions.
In conclusion, lattices with $T = L \pm a$ are interpreted as having physical time extent $T'$ with $T' = T + sa$ with $s =  \pm 1$, and it is then possible to set $T'= L$. This modification affects even the pure gauge theory,
so that we have to revisit the O($a$) effects there before turning
to the fermionic contributions.

\section{Pure gauge theory}

On the lattice we choose the usual Wilson plaquette gauge action,
\be\label{action}
S[U] = \frac{1}{g_0^2}\sum_p w(p){\rm tr }\left\{1 - U(p) \right\},
\ee
where the sum runs over all oriented plaquettes $p$, and $U(p)$ denotes the parallel transporter around $p$. The coefficients $w(p)$ are weight factors to be specified shortly. Due to the abelian nature of the boundary fields $C$ and $C'$ the lattice version of Eqs.~(\ref{bc_cont}) for the 
link variables $U(x,\mu)$ reads
\be
    U(0,{\bf x},k)=\exp(aC^{}_k),\qquad  U(T,{\bf x},k)=\exp(aC'_k).
\ee
With these boundary conditions it is known that the O($a$) effects
generated by the presence of the boundary are encoded in a single
operator,
\be
 a\int_{x_0 = 0,T } d^3 {\bf x} {\rm tr }\left\{F_{0k}F_{0k}\right\}.
 \label{boundary_operator}
\ee
An ${\rm O}(a)$  improved lattice action can thus be obtained from Eq.~(\ref{action}), by setting $w(p)$ equal to 1 except for the time-like plaquettes attached to one of the time boundaries, where one sets $w(P)=c_t(g_0)$. The coefficient $c_t$ then multiplies
a lattice version of (\ref{boundary_operator}), and, if chosen appropriately,
the O($a$) effects in observables are cancelled.

\subsection{Tree-level considerations}

In perturbation theory, $c_t$ is expanded
\be
c_t(g_0) = c_t^{(0)} +c_t^{(1)}g_0^2 + {\rm O}(g_0^4).
\ee
In the standard framework with $T=L$ one has $c_t^{(0)}=1$
and the next two coefficients are known, too~\cite{Bode:1999sm}.
However, when taking the continuum limit at fixed $T'/L = 1$, 
a modification of the tree-level coefficient $c_t^{(0)}$ is required.
To calculate it, we first need to determine the minimal action
configuration as a function of $c_t^{(0)}$.

\subsection{Equations of motion}

In order to be able to write the equations of motion concisely, 
we follow \cite{Luscher:1992an} and introduce the covariant divergence of 
the plaquette, slightly modified by the inclusion of the weight factors, 
\ba
d_w^*P(x,\mu) &=& \sum_{\nu=0}^3\left\{w\lbrack P_{\mu\nu}(x)\rbrack P_{\mu\nu}(x) \right. \nonumber\\&-&  \left. w\lbrack P_{\mu\nu}(x - \hat\nu)\rbrack U^\dagger(x - \hat\nu, \nu)P_{\mu\nu}(x -  \hat \nu) U(x - \hat \nu, \nu)\right\}.
\ea
The lattice action will be stationary if and only if the traceless antihermitian part of $d^*_wP(x, \mu)$ vanishes,
\be\label{movto}
d_w^*P(x,\mu) - d^*_wP^\dagger(x,\mu) -\frac 1N \textrm{tr }\left\{d_w^*P(x, \mu) - d^*_wP^\dagger(x, \mu)\right\} = 0.
\ee
We make an ansatz of the form,
\be
V(x,\mu) = \exp\left(aB_{\mu}(x_0)\right),
\ee
with a spatially constant and Abelian field $B_\mu(x_0)$.
Up to gauge equivalence, the equations of motion are then 
solved by, 
\be
B_0(x_0) = 0, \qquad 
B_k(x_0) = \left\{\begin{array}{ll} \left(x _ 0 - \frac T2\right)f + \frac{C_k + C_k'}{2} & x_0 \in \lbrack a, T-a \rbrack \\  C_k & x_0 = 0\\  C_k' & x_0 = T\end{array}\right.\\
\ee
where $f$ can be computed either numerically or as a power series
in $a$. To check whether the above ansatz really leads to the
absolute minimum gauge configuration, a cooling procedure has been
applied to random gauge configurations. While this does not constitute 
a proof, the apparent absence of configurations with lower action 
for various lattice sizes lends further support to our assumptions.

\subsection{Choice of $c_t^{(0)}$}

The tree-level coefficient $c_t^{(0)}$ is to be chosen such that ${\rm O}(a)$ effects in observables are cancelled. To this order we may just require the 
lattice action to coincide with its continuum counterpart up to O($a^2$) terms.
Expanding the lattice action to O($a$)
\be
S_{latt}=  S_{cont} \left\{ 1 + \frac aL  \left\lbrack -2 + s + \frac{4c_t^{(0)} - 2}{c_t^{(0)}}\right\rbrack +{\rm O}(a^2)\right\}.
\ee
we thus conclude
\be
c_t^{(0)} = \frac{2}{2 + s}. 
\ee
A closer look then reveals that this choice even removes the lattice artefacts in the action up to order $a^4$.

\subsection{One-loop calculation}

Working in a renormalisable gauge, the first two terms in the effective action, Eq.~(\ref{Effaction}),  
are given by
\be
\Gamma_0[B] = g_0^2 S[V],  \qquad\Gamma_1[B] = \frac 12 \ln \det \Delta_1 - \ln \det \Delta_0,
\ee
where $\Delta_0$ and $\Delta_1$ are the fluctuation operators  for the ghost fields and gauge fields respectively. 
The SF coupling to this order then becomes
\be
\bar g^2(L) = g_0^2 + m_1(L/a) g_0^4  + \ldots, \qquad \quad m_1(L/a) = - \Gamma'_1/\Gamma'_0 .
\ee
To compute the quantities,
\be
 \partial_\eta (\ln \det \Delta_j )\big/ \Gamma'_0, \qquad j = 0,1,
\ee
we have followed the strategy used in \cite{Luscher:1992an}.
One expects that $m_1(L/a)$ has an asymptotic expansion of the form,
\be
m_1(L/a) \stackrel{a/L \rightarrow 0}{\sim} \sum_{n= 0}^\infty(a/L)^n(B_n + A_n \ln (L/a)).
\ee
The results obtained for $m_1(L)$ have been confirmed by 
an independent calculation performed by S.~Takeda and U.~Wolff \cite{octavo}. 
The (preliminary) results obtained for $B_0$ and $B_1$ are shown in Table \ref{m1}, where we have set $A_0$ and $A_1$ to their expected values 
(after having confirmed them numerically).
\begin{table}[h!]
\begin{center}
\begin{tabular}{|c|c|c|c|c|}
\hline
 $s$ & $A_0$ & $B_0$ & $A_1$ & $B_1$ \\\cline{1-5} 
 $-1$ & $22/(4\pi)^2$ & $0.368283(1)$  & $0$ & $-0.2318(3)$ \\\cline{1-5} 
 $0$  & $22/(4\pi)^2$ & $0.3682817(7)$ & $0$ & $-0.1779(3)$\\\cline{1-5} 
 $1$  & $22/(4\pi)^2$ & $0.3682818(7)$ & $0$ & $0.1232(4)$ \\ 
\hline
\end{tabular}
\caption{Preliminary results for the coefficients in the asymptotic expansion (in collaboration with S.~Takeda and U.~Wolff).\label{m1}}
\end{center}
\end{table}

\subsection{Determination of $c_t^{(1)}$}

To determine the one-loop coefficient, we expand the lattice action as a Taylor series about $c_t = c_t^{(0)}$,  
\be
S_{latt} = S_{latt}\big|_{c_t^{} = c_t^{(0)}} + g_0^2 \frac{\partial S_{latt}}{\partial c_t}\big|_{c_t^{} = c_t^{(0)}} c_t^{(1)}.
\ee
Inserting this expansion in the definition of the coupling we arrive at, 
\be
\bar g^2(L) = g_0^2 + \left( m_1(L/a) - c_t^{(1)} \partial_{c_t}\Gamma'_0\big |_{c_t = c_t^{(0)}}\big/\Gamma_0'\big |_{c_t =c_t^{(0)}}\right)g_0^4 + {\rm O}(g_0^6)
\ee
The factor multiplying $c_t^{(1)}$ behaves like ${{\rm O}(a/L)}$, and can thus remove the contribution of $B_1$, if we adjust $c_t^{(1)}$ properly. We obtain
\be
c_t^{(1)} = \frac{B_1}{2} \left(c_t^{(0)}\right)^2. 
\ee

\section{Staggered fermion action}

The fermionic part of the action takes the form,
\be
S_{f} = a^4\sum_{\bf x}\sum_{x_0=a}^{T-a} \frac 1{2a} \eta_\mu (x)\bar \chi(x)\left\lbrack\lambda_\mu U(x,\mu) \chi(x + \hat\mu) - \lambda_\mu^{\dagger}U^{\dagger}(x - \hat\mu,\mu) \chi(x - \hat\mu) \right \rbrack  + S_B^{(0)} + S_B^{(T)},
\ee
where the last two terms encode fermionic boundary terms~\cite{Heller:1997pn}.
The coefficient $c_t^{(1)}$ also receives a contribution 
from the fermionic part of the action. We have obtained preliminary results for the contribution to $m_1(L/a)$; in particular, we find that $A_0$ and $B_0$ coincide with the results obtained by Heller~\cite{Heller:1997pn}. However, before we can quote a value for the staggered one-loop contribution to $c_t^{(1)}$, a more detailed analysis of the fermionic O($a$) boundary counterterm needs to be carried out.

\section{Conclusions}

First steps have been taken in a study of the SF running coupling 
in QCD with four quark flavours. Staggered fermions are a natural choice but require some revision of the standard framework, due to the fact that $L=T$ can only be imposed up to O($a$) terms. Our proposal to take the continuum limit 
at fixed $T' = L$ with $T' = T \pm a$ modifies the ${\rm O}(a)$ improvement counterterm proportional to $c_t$, which we then determined perturbatively at the tree and the one-loop level. As a byproduct this yields an alternative definition of the pure gauge SF which is currently being tested in simulations 
(in collaboration with S.~Takeda and U.~Wolff).
In the near future, this will hopefully be followed by numerical simulations including staggered fermions.

\section*{Acknowledgments}

A pleasant collaboration with S. Takeda and U. Wolff is gratefully acknowledged.

\end{document}